# CTLESS: A scatter-window projection and deep learning-based transmission-less attenuation compensation method for myocardial perfusion SPECT

Zitong Yu, Md Ashequr Rahman, Craig K. Abbey, Richard Laforest, Nancy A. Obuchowski, Barry A. Siegel, and Abhinav K. Jha

*Abstract*— Attenuation compensation (AC), while being beneficial for visual-interpretation tasks in myocardial perfusion imaging (MPI) by single-photon emission computed tomography (SPECT), typically requires the availability of a separate X-ray CT component, leading to additional radiation dose, higher costs, and potentially inaccurate diagnosis in case of misalignment between SPECT and CT images. To address these issues, we developed a method for <u>c</u>ardiac SPEC<u>T</u> AC using deep <u>l</u>earning and <u>e</u>mission <u>s</u>catter-window photon<u>s</u> without a separate transmission scan (CTLESS). In this method, an estimated attenuation map reconstructed from scatter-energy window projections is segmented into different regions using a multi-channel input multi-decoder network trained on CT scans. Predefined attenuation coefficients are assigned to these regions, yielding the attenuation map used for AC. We objectively evaluated this method in a retrospective study with anonymized clinical SPECT/CT stress MPI images on the clinical task of detecting perfusion defects with an anthropomorphic model observer. CTLESS yielded statistically non-inferior performance compared to a CT-based AC (CTAC) method and significantly outperformed a non- AC (NAC) method on this clinical task. Similar results were observed in stratified analyses with different sexes, defect extents, and defect severities. The method was observed to generalize across two SPECT scanners, each with a different camera. In addition, CTLESS yielded similar performance as CTAC and outperformed NAC method on the fidelity-based figures of merit, namely, root mean squared error (RMSE) and structural similarity index measure (SSIM). Moreover, as we reduced the training dataset size, CTLESS yielded relatively stable AUC values and generally outperformed another DL-based AC method that directly estimated the attenuation coefficient within each voxel. These results demonstrate the capability of the CT- LESS method for transmission-less AC in SPECT and motivate further clinical evaluation.

This work was supported in part by National Institute of Biomedical Imaging and Bioengineering of National Institute of Health (NIH) under grants R01-EB031051, and R01-EB031962, Bradley-Alavi Student Fellowship from the Education and Research Foundation for Nuclear Medicine and Molecular Imaging (ERF) of the SNMMI, and the National Science Foundation (NSF) CAREER Award 2239707.

Zitong Yu is with Department of Biomedical Engineering, Washington University, St. Louis, MO, USA (e-mail: yu.zitong@wustl.edu).

Md Ashequr Rahman is with Department of Biomedical Engineering, Washington University, St. Louis, MO, USA (e-mail: rahman.m@wustl.edu).

Craig K. Abbey is with Department of Psychological & Brain Sciences, University of California Santa Barbara, Santa Barbara, CA, USA (e-mail: ckabbey@ucsb.edu).

Nancy A. Obuchowski is with Quantitative Health Sciences, Lerner Research Institute, Cleveland Clinic, Cleveland, OH, USA (e-mail: nobuchow@ccf.org).

Richard Laforest is with Mallinckrodt Institute of Radiology, Washington University, St. Louis, MO, USA (e-mail: rlaforest@wustl.edu).

Barry A. Siegel is with Mallinckrodt Institute of Radiology, Washington University, St. Louis, MO, USA (e-mail: siegelb@wustl.edu).

Abhinav K. Jha is with Department of Biomedical Engineering and Mallinckrodt Institute of Radiology, Washington University, St. Louis, MO, USA (e-mail: a.jha@wustl.edu).

## I. INTRODUCTION

ATTENUATION of photons is a major image-quality-degrading effect in single-photon emission computed tomography (SPECT) myocardial perfusion imaging (MPI). Multiple studies have shown that attenuation compensation (AC) is beneficial for clinical interpretations of MPI-SPECT images [1], [2]. AC methods require an attenuation map, which is obtained with a transmission source in earlier SPECT scanners or with an additional CT transmission scan in modern SPECT/CT cameras [3], [4]. However, this has multiple disadvantages, including higher costs, slightly increased radiation dose, and potential for inaccurate diagnosis due to possible misregistration between SPECT and CT images [5], [6]. Further, many SPECT systems do not have a CT component. These include SPECT systems in many community hospitals and most physician offices, mobile SPECT systems that enable imaging in remote locations, and the emerging solid-state- detector-based SPECT systems that otherwise provide higher sensitivity, as well as higher energy, temporal and spatial resolution compared to conventional SPECT systems [7], [8]. Due to all these reasons, approximately 74% of MPI-SPECT studies are performed without AC worldwide [9]. Thus, there is an important need to develop transmission-less AC (Tx-less AC) methods for SPECT.

Given this need, multiple Tx-less AC methods have been proposed. One set of methods focused on using the physics of the SPECT emission data [10]–[15]. These methods can be divided into two categories. The first category of methods estimates attenuation coefficients directly from the SPECT emission data. These methods operate on iterative inversion of the forward mathematical models of SPECT systems [10], [11], or use the consistency conditions grounded in the forward model [12]. The second category of methods use the SPECT scatter-window data to estimate attenuation maps. This includes approaches that obtain an attenuation map by assigning attenuation coefficients to segments of an initial estimate of the attenuation map [14] and approaches based on inversion of scattering models [15]. However, these physics-based methods have limited accuracy and have met with only limited success [6].

More recently, deep learning (DL)-based methods have been proposed for AC in SPECT [16]–[21]. These can be divided into direct and indirect methods [18]. The direct methods directly estimate activity maps with AC from non-AC activity maps [19], [20] while the indirect methods estimate attenuation maps from non-AC activity maps [21]. These methods have shown promise, but they seem to either rely on correlation between the non-AC and AC SPECT activity maps for direct methods or the relation between the non- AC SPECT activity map and the attenuation map for indirect methods. The premise for this correlation remains unclear so the physical foundation of the proposed methods remains limited. In clinical MPI-SPECT acquisition with most modern cameras, projection data are collected in both photopeak and scatter-energy windows. In this context, we recognize that studies have shown that scatter-window projection data contains information to estimate the attenuation distribution [22]. Further, since the probability of scatter at a certain location is proportional to the attenuation distribution at that location, it is expected that a reconstruction of the scatter-energy window projection would exhibit contrast between regions with different attenuation coefficients. Previously proposed physics-based methods have used the idea of segmenting the scatter-window reconstruction into different regions, to which pre-defined attenuation coefficients are assigned to yield an estimate of the attenuation map [10], [11], [14]. A previous study by Pan et al. has shown that an activity map reconstructed using the photopeak-energy window projection could assist with such segmentation task by providing additional anatomic information. [23]. However, these segmentations have limited accuracy. In this context, with cardiac SPECT, large amounts of SPECT emission data and corresponding CT scans are avail- able, where the CT scans can be segmented to obtain ground- truth segmentations. Thus, we investigated the use of deep learning to segment the initial scatter-window reconstruction with photopeak-window reconstruction providing additional assistance. Building on the idea of integrating the physics and DL-based methods, we propose a method for cardiac SPECT AC using deep learning and emission scatter-window photons without a separate transmission scan (CTLESS).

Another limitation of existing DL-based AC methods is that they have typically been evaluated using figures of merit (FoMs) that measure the fidelity between the images reconstructed using the DL-based approach with a reference standard [24]–[26]. However, studies have shown that evaluation using such FoMs may not correlate with performance on clinical tasks in MPI [27], [28]. Thus, it is necessary to evaluate these methods on the specific clinical task for which the images are acquired. The need for task-based evaluation was also recommended in the recently proposed best practices for evaluation of AI algorithms for nuclear medicine (the RELAINCE guidelines) [29]. Thus, in addition to conventional evaluation, we also objectively evaluate CTLESS on the clinical task of detecting myocardial perfusion defects in a retrospective study with anonymized clinical data from patients who underwent MPI SPECT studies.

## II. METHODS

### A. Theory

#### 1) Problem formulation

Consider a SPECT system imaging a tracer distribution within a human body, denoted by a vector $f(\mathbf{r})$, where $\mathbf{r} \in \mathbb{R}^3$ denotes the 3-dimensional coordinates. The SPECT system yields the projection data both in photo- peak and scatter energy windows, denoted by $M$-dimensional vectors $\boldsymbol{g}_{pp}$ and $\boldsymbol{g}_{sc}$, respectively, where $M$ is the number of elements in the SPECT projection.

The goal in SPECT is to reconstruct the tracer distribution given the projection data. To perform AC during reconstruction, an attenuation map is needed. The attenuation map is denoted by an $N$-dimensional vector $\boldsymbol{\mu}$, where $N$ is the number of voxels in the SPECT reconstructed image. Denoting the reconstructed image by an $N$-dimensional vector $\hat{\boldsymbol{f}}$, and the reconstruction operator that performs the AC by $\mathcal{R}_{\boldsymbol{\mu}}$, we have

$$\hat{\boldsymbol{f}} = \mathcal{R}_{\boldsymbol{\mu}}(\boldsymbol{g}_{pp}). \tag{1}$$

Conventional AC methods obtain the attenuation map $\boldsymbol{\mu}$ from a separate CT scan or a transmission-source scan [3], [4]. This has several issues, as mentioned in the introduction. Our goal is to estimate $\boldsymbol{\mu}$ only using the SPECT emission data $\boldsymbol{g}$.

#### 2) Proposed CTLESS method

As mentioned earlier, we investigated the use of SPECT emission data $\boldsymbol{g}_{pp}$ and $\boldsymbol{g}_{sc}$ for estimating an attenuation map, followed by using DL to further refine this attenuation map. We mathematically frame the problem as segmenting the initial scatter-window reconstruction into $K$ attenuation regions. The CT scans are segmented to provide a surrogate for the ground-truth segmentation. Consider that the CT scan is segmented into $K$ attenuation regions, each corresponding to an organ. To denote the support for the segmented kth attenuation region, we define an $N$-dimensional vector $\boldsymbol{\Phi}^k$, the $i^{th}$ element of which is given by

$$\boldsymbol{\Phi}_i^k = \begin{cases} 1, & \text{if } i^{th} \text{ voxel in } k^{th} \text{ region} \\ 0, & \text{otherwise} \end{cases}. \tag{2}$$

Both photopeak and scatter-window reconstructions are used to perform the segmentation operation. The scatter window projection $\boldsymbol{g}_{sc}$ is reconstructed using an ordered-subsets expectation maximization (OSEM) technique that compensates for collimator-detector response and Poisson noise in the scatter-window data. The scatter-window reconstruction $\hat{f}_{sc}$ is given by Eq. 3, where $\mathcal{R}$ is the reconstruction operator without AC.

$$\hat{f}_{sc} = \mathcal{R}(g_{sc}). \tag{3}$$

Similarly, the photopeak-window reconstruction, denoted by $\hat{f}_{pp}$, is reconstructed using the same OSEM technique as follows.

$$\hat{f}_{pp} = \mathcal{R}(g_{pp}). \tag{4}$$

Consider a segmentation operator parameterized by the parameter vector $\boldsymbol{\theta}$, and denoted by $\mathcal{D}_{\boldsymbol{\theta}}$. Both $\hat{f}_{sc}$ and $\hat{f}_{pp}$ are the input of the segmentation operator by $\mathcal{D}_{\boldsymbol{\theta}}$, which is trained to yield the estimated segmentation map, i.e.,

$$\{\hat{\boldsymbol{\Phi}}\} = \mathcal{D}_{\boldsymbol{\theta}}(\hat{f}_{sc}, \hat{f}_{pp}), \tag{5}$$

where $\{\hat{\boldsymbol{\Phi}}\} = \{\hat{\boldsymbol{\Phi}}^1, \hat{\boldsymbol{\Phi}}^2, ..., \hat{\boldsymbol{\Phi}}^K\}$ is the set of segments estimated by the segmentation operator $\mathcal{D}_{\boldsymbol{\theta}}$.

We developed a multi-channel input and multi-decoder U-net (McEUN) to realize this operator DΘ. The McEUN consists of two components: an encoder with multi-channel input and an assembly of decoders, where the number of decoders equals the number of regions to be segmented. The input of the encoder consists of two channels: one channel receives the 3-D scatter-window reconstructed images, and the other channel receives the 3-D photopeak-window reconstructed images. The encoder contains five convolutional layers with 3×3×3 kernel, decreasing the size of input images from 64 × 64 × 64 to 16 × 16 × 16. This reduction in image size was achieved by two of these convolutional layers, each with a stride size of 2. The numbers of filters in convolutional layers are doubled as the size of image decreases each time. Each decoder contains three blocks of a transposed convolution with 3 × 3 × 3 kernel and a convolution with 3×3×3 kernel. The numbers of filters are halved as the size of image increases each time. After each convolutional layer in both the encoder and decoders, a leaky rectified linear unit is applied. To stabilize the network training and improve segmentation performance, skip connections with attention gate are provided between the encoder and each decoder [30]. The encoder extracts local spatial features from initial estimates of activity and attenuation maps. The decoders map the extracted features to the segmentation of specific regions. In the final layer, the outputs of six decoders are concatenated and a SoftMax function is applied, yielding the final segmentation of the entire image. The architecture of McEUN is shown in Fig. 1 with detailed descriptions in the supplementary materials.

The segmentation network was optimized to minimize the weighted cross-entropy between the predicted segmentations and CT-derived segmentations. The loss function for each image, denoted by $\mathcal{L}(\boldsymbol{\theta})$, is given by

$$\mathcal{L}(\boldsymbol{\theta}) = \sum_{i=1}^{N}\sum_{k=1}^{K} w_k \left[ -\Phi_i^k \log \hat{\Phi}_i^k - (1-\Phi_i^k)\log(1-\hat{\Phi}_i^k) \right], \tag{6}$$

where $w_k$ is the weight parameter for the $k^{th}$ region. By optimizing this loss, the McEUN was trained to yield the estimated segments $\{\hat{\boldsymbol{\Phi}}\}$. We assume that the attenuation coefficient in each region is constant, denoted by a scalar value $\mu_k$, where $k$ is the region index, and assigned to corresponding estimated region segments. Thus, the final estimated attenuation map, denoted by an $N$-dimensional vector $\hat{\boldsymbol{\mu}}$, is given by

$$\hat{\boldsymbol{\mu}} = \sum_{k=1}^{K} \mu_k \hat{\boldsymbol{\Phi}}^k. \tag{7}$$

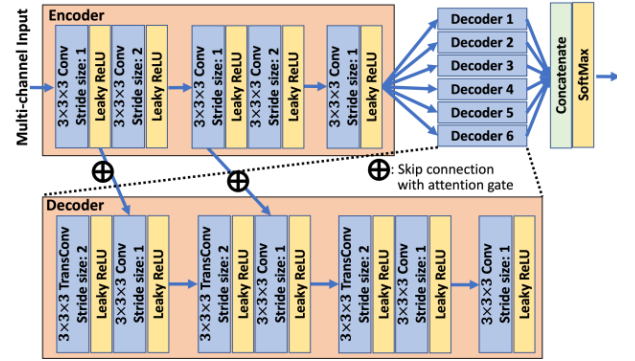

Fig. 1. The architecture of the multi-channel input and multi-encoder U-net.

The final estimated attenuation map is then used for AC in the reconstruction as in Eq. (1). The overall framework of this approach, which we refer to as CTLESS, is shown in Fig. 2.

### B. Evaluation

We quantitatively evaluated the performance of CTLESS on the cardiac defect-detection task in an IRB-approved retrospective clinical study (IRB ID 201905164). To maintain high rigor in our evaluation, we followed the RELAINCE guidelines [29]. Our reference standard for this evaluation study was activity maps reconstructed using a CT-based AC (CTAC) method. The CTLESS method was also compared to a method where no AC was performed, referred to as no- AC (NAC) approach. Stratified analyses were conducted for different sexes, defect extents, and defect severities. Further, the generalizability of CTLESS to two different scanners was evaluated. We also quantitatively assessed the visual similarity of the images yielded using CTLESS with the reference standard. Finally, we assessed the performance of CTLESS with different sizes of the training dataset, where we compared the method with another DL-based AC approach that directly estimates the attenuation coefficients of the different voxels in the attenuation map, similar to previously proposed approaches [20], [21].

The evaluation study consisted of four parts, including (1) data collection, curation, and power analysis, (2) network training and evaluation with test data, (3) process to extract task-specific information, and (4) figures of merit and statistical analyses.

#### 1) Data collection, curation, and power analysis

A flowchart of the data collection and curation process in our study is shown in Fig. 3.

*Data Collection:* A database of N=3719 anonymized patients who had undergone rest and stress MPI studies was considered.

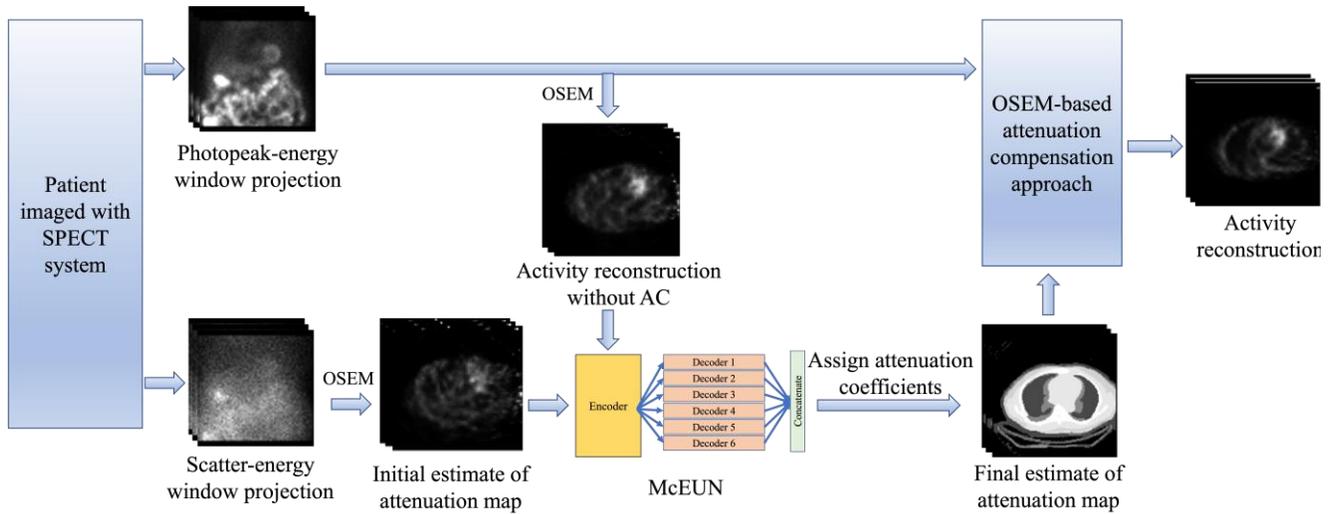

Fig. 2. The overall framework of the CTLESS method.

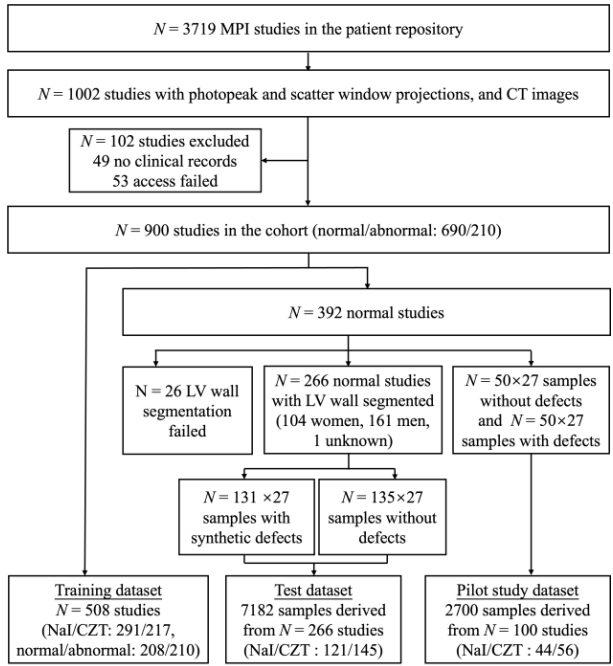

Fig. 3. Flow chart of the data collection and curation process in our retrospective evaluation study.

The clinical protocol was a one-day stress/rest protocol and the mean injected activity for the stress images was 10 mCi in patients weighing under 250 pounds and 12 mCi for those over 250 pounds. 1002 MPI studies contained SPECT projection data in photopeak and scatter windows and CT images. We obtained N = 900 stress MPI studies along with patient sex information and clinical reports. Based on clinical reports, we categorized studies with normal myocardial perfusion images as normal (N = 690), while studies with perfusion defects in the left ventricular wall were categorized as abnormal (N = 210). MPI scans were acquired on two SPECT/CT scanners: GE Discovery NM/CT 670 Pro Sodium- Iodide (NaI) (NaI-camera) (N = 456) and GE Discovery NM/CT 670 Pro CZT (CZT-camera) (N = 418). N = 26 studies were excluded as we will mention later.

*Description of Image Acquisition and Processing Parameters*: On both scanners, SPECT emission data were collected in photopeak and scatter windows after the injection of 99mTc-tetrofosmin. Detailed SPECT image acquisition and reconstruction parameters are listed in Table I. CT images were acquired at 120 kVp source energy at 10 mA tube current, revolution time of 0.8 seconds, and a spiral pitch factor of 0.9375, with the GE Optima CT 540 system component of the scanner. The CT dose index-volume ($CTDI_{volume}$) was 9.1 mGy/100mAs. The CT images were collected at low dose and only used for AC.

The CT and SPECT images were registered using MIM Maestro (MIM Software Inc, Cleveland, OH). More specifically, the MIM Maestro software first reconstructed the SPECT data without AC and provided an initial registration. Then, we manually adjusted the registration through 3D translation, guided by feedback from our clinical collaborators. CT-defined attenuation maps were calculated from the CT scans using a bi-linear model [31]. The CTAC method reconstructed the photopeak-window projections using an OSEM-based method with 8 iterations and 6 subsets and CT-based attenuation maps for AC on both the NaI-camera and CZT-camera [32]. The OSEM-based reconstruction method compensated for the major image-degrading artifacts in SPECT, including attenuation and collimator-detector response. The reconstructed images had a size of $64 \times 64 \times 64$ with a voxel size of 0.68 cm. As per clinical protocol, these images were then filtered using a Butterworth filter with an order of 5 and cutoff frequency of 0.44 cycles per cm and reoriented into short-axis slices.

*Defining Defects*: To evaluate the performance of the CTLESS method on the defect-detection task, the knowledge of the existence and location of the defects in the defect-present test images was needed. To ensure precise and unbiased assessment, we used only normal studies in the test set and inserted synthetic cardiac defects in images in half of these normal studies to create the defect-present population. The remaining half of the normal studies were referred to as defect-absent population. Following a similar procedure as Narayanan et al. [33], we inserted 27 types of realistic defects into the photopeak window projection with three radial extents (30, 60, and 90 degrees around the left ventricular (LV) wall), three severities (10%, 25%, and 50% less activity than the normal myocardium), and at three locations (anterior, inferior, and lateral walls of the LV). The synthetic defects were similar to the characteristics of the defects considered in the previous studies [34], [35]. We did not insert these synthetic defects into the scatter window projections. Therefore, the scatter window projections were kept the same. N = 26 studies were excluded due to LV wall segmentation failure, since LV wall segmentation was needed for the insertion of synthetic defects. Details of the defect insertion procedure were provided in the supplementary materials. Sample images of defect types are shown in Fig. 4.

TABLE I: The acquisition and reconstruction parameters used on NaI-camera and CZT-camera

| SPECT scanner | NaI-camera | CZT-camera |
|---|---|---|
| Collimator type | Low energy high resolution | Wide energy high resolution |
| Collimator grid | Parallel hole | |
| Energy resolution (@140 keV) | 9.8% | 6.3% |
| System resolution (FWHM@100 mm) | 7.4 mm | 7.6 mm |
| System sensitivity (@100 mm per detector) | 72 cps/MBq | 85 cps/MBq |
| Number of views | 30 | |
| Orbit | 180° ranging from 45° right anterior oblique to 45° left posterior oblique | |
| Acquisition mode | Step and shoot | |
| Imaging time | Around 40 seconds per projection view | |
| Scatter-energy window | 114-126 keV | |
| Photopeak-energy window | 126-154 keV | |
| Reconstruction method | OSEM with 8 iterations and 6 subsets | |

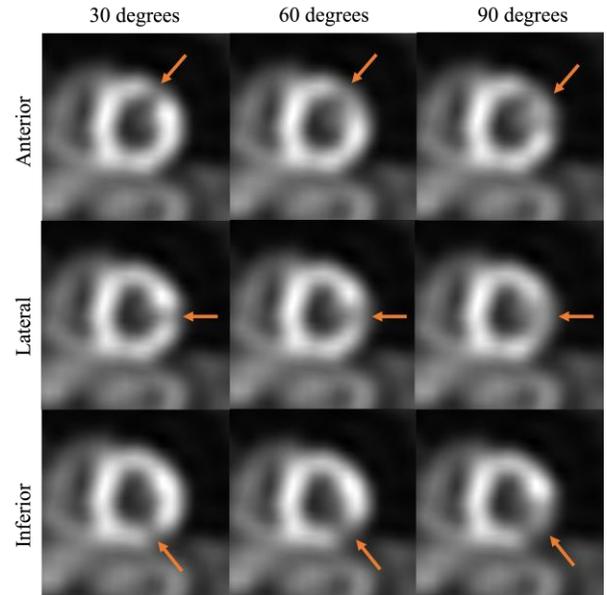

Fig. 4. Sample images of defect types from the short-axis view of the myocardium for different extents (along the rows) and at different locations (along the columns). The defect severity is set as 50% in these images for ease of visualization. Red arrows indicate defect locations.

*Power Analysis*: The primary objective of our study was to assess the non-inferiority of the CTLESS method to the CTAC method on cardiac defect-detection task, as measured by the area under the receiver operating characteristics curve (AUC). The non-inferiority margin was defined as 5% of the AUC obtained by the CTAC method, assuming a moderate observer variability [36]. To determine the size of test dataset, we conducted a power analysis using a strategy proposed by Obuchowski [37]. We first needed an estimate of the variance of the difference in AUCs between CTAC and CTLESS methods. To obtain these estimates, we conducted a pilot study with data from N = 100 normal patients. In half of these patients, 27 types of defects were introduced for a total of 27 × 50 = 1350 cases with defects. We also generated N = 1350 (50 × 27) defect-absent samples from the other half of the patients for a prevalence rate of 50%. Following a similar strategy as outlined in Sec. II-B.3 and Sec. II-B.4, we observed that in this pilot study, we obtained an estimate of the variance of the difference in AUC values between CTAC and CTLESS methods. Our power analysis revealed that for the considered non-inferiority margin, to show the non-inferiority of CTLESS to CTAC with a significance level of 0.05 and a power of 0.8, N = 129 defect-absent and N = 129 defect-present studies were needed. Based on this, we considered N = 266 independent MPI studies in the test dataset, of which 131 were defect-present and 135 were defect-absent.

Generation of Test Dataset: For the test dataset, we generated 27 × 131 = 3537 defect-present samples. For the remaining 135 MPI studies, we generated 27 × 135 = 3645 defect-absent samples, although samples from the same patient were identical. On both the NaI-camera and CZT-camera, the scatter-window projections were reconstructed using an OSEM-based method with the same parameters as in the CTAC method but without AC, yielding the initial estimate of attenuation maps. The photopeak-window projections were also reconstructed using the same strategy. Examples of photopeak-window and scatter-window reconstructions, as well as activity reconstruction obtained by CTAC method from both cameras are shown in Fig. 5. The CT-based attenuation maps used for training were segmented into skin and subcutaneous adipose, muscles and organs, lungs, bones, patient table, and background, using a Markov random field-based method [38]. The average attenuation coefficient of each region was calculated and served as the predefined attenuation coefficient values, as listed in Table II.

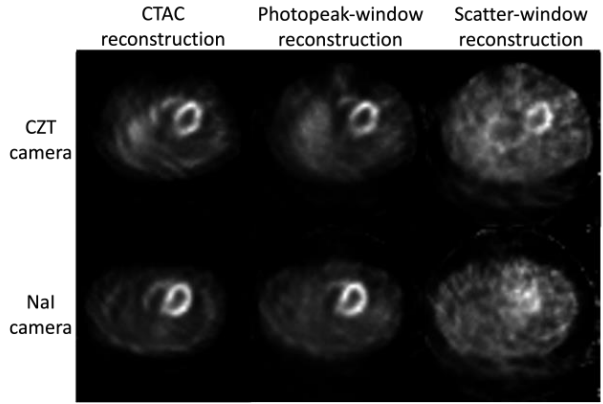

Fig. 5. Examples of photopeak-window reconstruction, scatter-window reconstructions, and activity reconstruction obtained by CTAC method on both the NaI-camera and CZT-camera.

TABLE II: Statistical parameters of attenuation coefficients across the patient population and predefined attenuation coefficients used in CTLESS

| Region | Attenuation coefficient value* (in cm$^{-1}$) | Pre-defined attenuation coefficient (in cm$^{-1}$) |
|---|---|---|
| Skin and subcutaneous adipose | 0.1339 (0.0029, 0.1337) | 0.13 |
| Muscles and organs | 0.1604 (0.0022, 0.1608) | 0.16 |
| Lungs | 0.0344 (0.0035, 0.0338) | 0.03 |
| Bones | 0.2156 (0.0048, 0.2159) | 0.22 |
| Patient table | 0.0884 (0.0045, 0.0878) | 0.09 |
| Background | 0.004 (0.005, 0) | 0 |

*2) Network training and evaluation with test data*

A total of 508 samples were used for network training. The initial estimates of attenuation maps and the photopeak-energy window reconstruction without AC were input to the McEUN. The network was trained to minimize the loss function as given by Eq. 6. To improve segmentation performance [21], the initial estimates of attenuation maps and the photopeak- window reconstruction without AC were normalized from [0, maximum pixel value in the image] to [0, 1] before being input to the network. The kernel weights of the McEUN were initialized using the Glorot normal initializer [39]. Biases were initialized to a constant of 0.03. Dropout with a rate of 0.1 was applied to prevent overfitting [40]. The McEUN was trained using Adam optimizer [41]. We optimized the weight parameters in Eq. (6) via five-fold cross validation. The optimized weight parameters are shown in the supplementary materials. The training and validation were performed using Keras 2.2.4 on two TITAN RTX GPUs, each with 24 GB of memory. The final training took around 2.5 hours. The number of epochs was chosen where the minimum averaged five-fold cross-validation loss was reached. The training and validation loss curves are shown in the supplementary materials. The segmentation network was designed to yield an output of 128×128×64 voxels. These dimensions were chosen since in our dataset, the CT images had the same dimensions, enabling comparison of the visual fidelity of the CTLESS-predicted and CT-derived attenuation maps using the metrics of root mean squared error (RMSE) and structural similarity index measure (SSIM).

The CTLESS method (Fig. 2) was used to generate the reconstructed activity image for the test dataset, using the same OSEM-based approach and post-processing procedures used in the CTAC method but with CTLESS-derived attenuation maps. NAC-based images were obtained using the same OSEM-based reconstruction approach and post-processing procedures but without AC.

*3) Process to extract task-specific information*

We objectively evaluated CTLESS on the task of detecting myocardial perfusion defects in a model-observer study. While ideally, such evaluation should be performed with human observers, this is time-consuming, tedious, and requires the availability of trained observers. At an early stage of translation, model observers provide a practical in silico approach to perform such evaluation and identify promising methods for subsequent evaluation with human observers [42].

In this study, the defect location was known, defect extents and severities were varying, and the background was varying. In previous MPI SPECT studies, for this defect-detection task with similar kinds of defects, it has been observed that the channelized Hotelling observer (CHO) with rotationally symmetric frequency (RSF) channels can emulate human observer performance [34], [35]. We thus used this model observer in this study. We used four RSF channels whose start frequency and width of the first channel were both 0.046 cycles per cm, as used in previous studies [34], [35], giving it a one-octave bandwidth. Subsequent channels were adjacent to the previous one with double the start-frequency and double the channel width, thereby maintaining octave bandwidth.

To apply this CHO, we extracted a $32 \times 32$ region (21.76 cm × 21.76 cm) from the middle 2-dimensional slice of the short-axis images such that the defect centroid was at the center of the extracted image, consistent with previous studies [34], [43]. For a better dynamic range in the cardiac region, we set the range of pixel values to [0, $x_{LV}$], where $x_{LV}$ is the maximum pixel value within the LV wall. Then, the pixel values were mapped to the range of [0, 255]. Denote the extracted images by $\hat{\boldsymbol{f}}_{SA}$, and the RSF channels by $\boldsymbol{U}$. We applied the RSF channels on the extracted images, yielding the feature vectors, denoted by $\boldsymbol{v}$, as follows:

$$\boldsymbol{v} = \boldsymbol{U}^T \hat{\boldsymbol{f}}_{SA}. \quad (8)$$

Denote the mean feature vectors obtained from defect-present and defect-absent samples by $\bar{\boldsymbol{v}}_s$ and $\bar{\boldsymbol{v}}_n$, respectively. We computed the mean difference in defect-present and defect-absent feature vectors, denoted by $\Delta \bar{\boldsymbol{v}}$, and given by

$$\Delta \bar{\boldsymbol{v}} = \bar{\boldsymbol{v}}_s - \bar{\boldsymbol{v}}_n. \quad (9)$$

Following this, the covariance matrix of the feature vectors, denoted by $\boldsymbol{K}_v$, was computed. These quantities were used to compute the template for the observer, denoted by $\boldsymbol{w}$, and given by

$$w = K_v^{-1} \Delta \overline{v}. \quad (10)$$

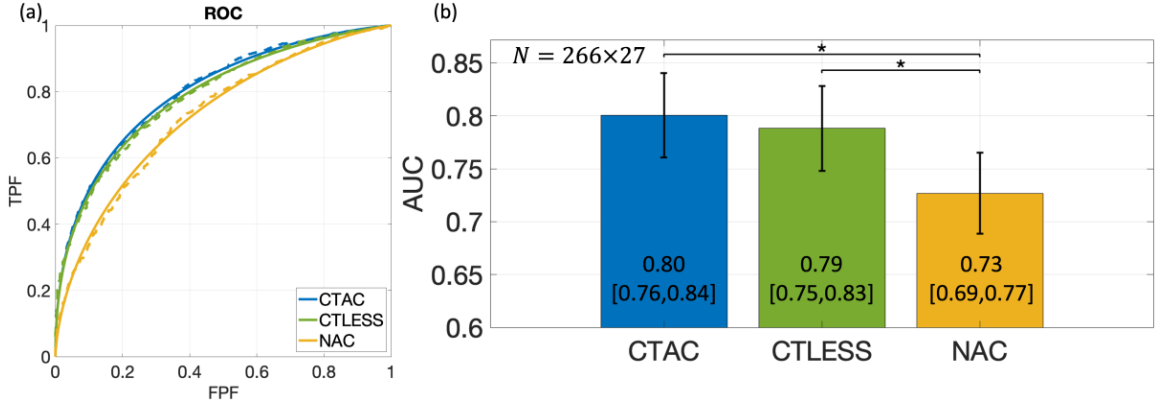

Fig. 6. (a) Fitted (solid lines) and empirical (dashed lines) ROC curves, and (b) AUC values obtained by the CTAC, CTLESS, and NAC methods on the perfusion defect-detection task from MPI-SPECT images. Stars indicate *p*-values less than 0.05.

The template of CHO was learned from the feature vectors of defect-present and defect-absent populations using a leave-one-out strategy. Applying this template to the feature vector for each sample in the testing dataset yielded a test statistic, denoted by $t$, given by

$$t = w^T v. \quad (11)$$

The test statistic was then compared to a threshold to classify each test sample into defect-present or defect-absent classes. By varying the threshold, we calculated the true-positive rate and false-positive rate and plotted a receiver operating characteristics (ROC) curve using LABROC4 program [44], [45].

### 4) Figures of merit and statistical analyses

We calculated AUCs with 95% confidence intervals (CIs) for CTLESS, CTAC, and NAC methods and differences in AUCs with 95% CIs among these methods, using a nonparametric strategy that accounted for the correlated nature of the data [37], [46], where the correlation between data samples was obtained from Hanley and McNeil [47]. Using these data, we first conducted a non-inferiority test to investigate whether the CTLESS method was at least as good as the CTAC method on the defect-detection task [48]. Next, we investigated whether CTLESS and CTAC were superior to NAC on the defect-detection task. To account for multiple hypothesis testing (CTLESS vs. NAC, and CTAC vs. NAC), Bonferroni correction was applied.

We also conducted stratified analyses for male and female populations, as well as for different defect extents and severity levels. In addition, we evaluated the generalizability of the CTLESS method across the two different scanners. We trained the CTLESS method using data acquired from the NaI-camera and tested on the data acquired from the CZT-camera, and vice versa.

Further, to assess the visual similarity of CTLESS with images yielded using the CTAC method, we computed the RMSE and SSIM between the images yielded with CTLESS and those from CTAC. We assessed statistical significance using a Bootstrap-based approach.

Finally, we assessed the performance of CTLESS with different sizes of the training dataset. The CTLESS method was compared with another DL-based AC approach that directly estimates attenuation coefficients within voxels (EST-AC), similar to previously proposed methods [20], [21]. Details of the EST-AC method are provided in the supplemental material. The size of the training dataset was varied from 26 to 508 patients. Both CTLESS and EST-AC were evaluated on the same test set with N = 266 patients. The performance was quantitatively compared on both the defect-detection task as well as the metrics of RMSE and SSIM.

For all statistical tests in this study, a *p*-value < 0.05 was used to infer statistical significance.

## III. RESULTS

### A. Evaluation on the cardiac defect-detection task

Fig. 6a shows ROC curves obtained by CTLESS, CTAC, and NAC methods. We found that the ROC curve obtained by the CTLESS method almost overlapped that obtained by the CTAC method and outperformed that of the NAC method. Fig. 6b shows the AUC values with CIs obtained by the three methods. We observed that the CTLESS method yielded a significantly higher AUC than that obtained by the NAC method. In the non-inferiority test, the lower limit of CIs of the AUC difference between the CTLESS and the CTAC method was within a margin of 5% of CTAC-based AUC (Fig. 7). Thus, the performance of CTLESS was deemed to be statistically non-inferior to the CTAC method within the predefined margin [48].

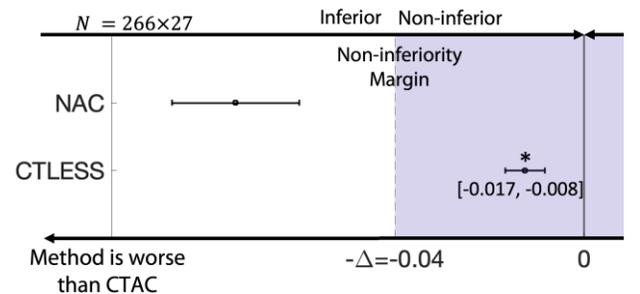

Fig. 7. Assessment of whether CTLESS is non-inferior to CTAC. NAC was treated as a placebo. The margin, denoted by Δ, was set to be 5% of CTAC-obtained AUC. Horizontal solid black lines indicate 95% confidence intervals of difference in AUC. Stars indicate *p*-values less than 0.05.

Fig. 8 shows the AUC values obtained with female and male subjects. For both sexes, the AUCs of the CTLESS method were significantly higher than those of the NAC method and similar to those of the CTAC method. Fig. 9 shows the AUC values as a function of defect extent and severity. The AUCs of the CTLESS method were close to those of the CTAC method for all considered defect extents and severity.

Fig. 10 shows results on the generalizability of the CTLESS method across two scanners. We observed that when trained on data acquired from the NaI-camera and evaluated on data from the CZT-camera, the CTLESS method had similar performance as the CTAC method on the defect-detection task. Similar results were observed when the CTLESS method was trained on data acquired from the CZT-camera and evaluated on data from the NaI-camera. When the CTLESS method was trained and evaluated on the same single scanner data, we observed that the performance was not significantly better than when it was trained on the dataset from both scanners. Further, in all training and evaluation settings, the CTLESS method significantly outperformed the NAC method.

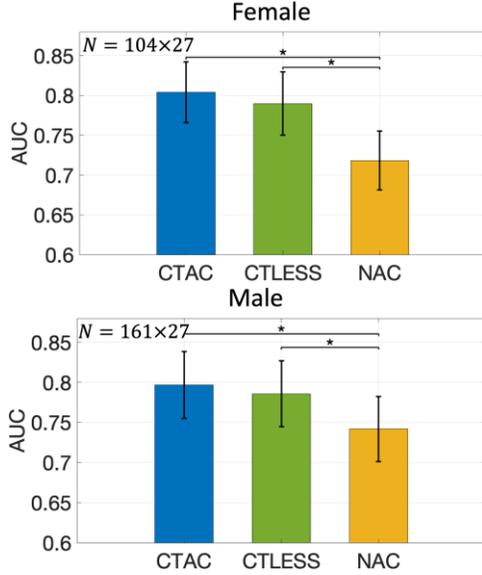

Fig. 8. AUC obtained by CTAC, CTLESS, and NAC methods on defect-detection task from MPI-SPECT images with different sexes. Stars indicate *p*-values less than 0.05.

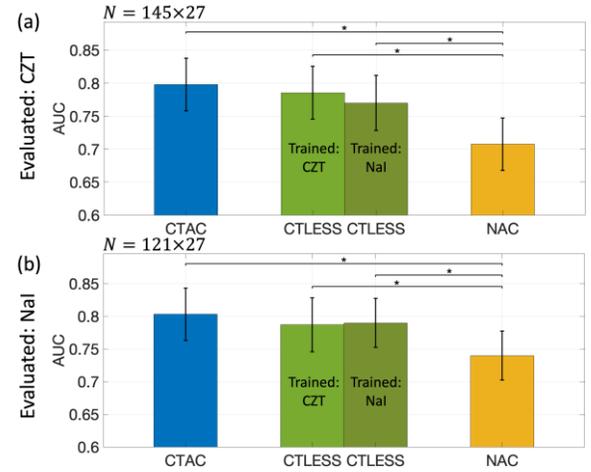

Fig. 10. AUC obtained by CTAC, CTLESS, and NAC methods across scanners. (a) AC methods were evaluated on data acquired from the CZT-camera. (b) AC methods were evaluated on data acquired from the NaI-camera. Stars indicate *p*-values less than 0.05.

### B. Evaluation based on fidelity-based figures of merit and representative examples

The evaluation study using the metrics of RMSE and SSIM showed that the CTLESS method significantly outperformed the NAC method (Fig. 11). Fig. 12 shows two representative examples of SPECT images and corresponding attenuation maps obtained using the CTAC, CTLESS, and NAC methods. We observe that the activity map obtained using CTLESS looked visually similar to those obtained using CTAC. Similar observations were found between the attenuation map obtained using CTLESS and CT-derived attenuation maps. Further, Fig. 12b shows a representative example of a defect-present case, where the defect was similarly observed in both the CTAC and CTLESS-obtained images. In contrast, activity images obtained using the NAC method were observed to have false-positive defects in the inferior LV wall in both examples.

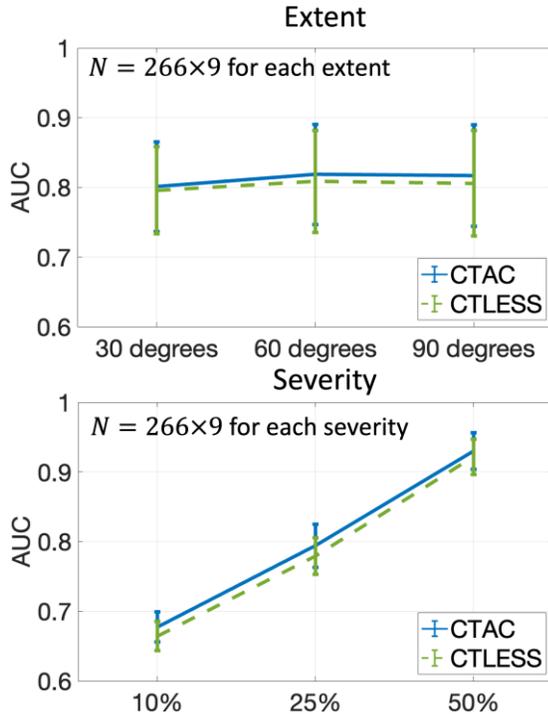

Fig. 9. AUC obtained by CTAC and CTLESS methods on the defect-detection task from MPI-SPECT images with different extents and severity levels of the defect.

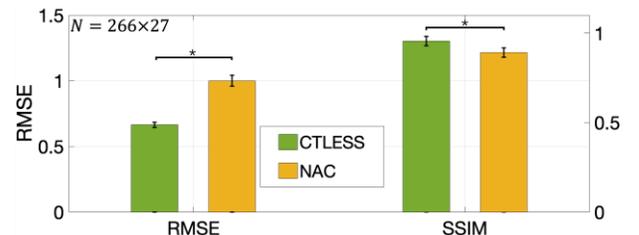

Fig. 11. RMSE and SSIM obtained using the CTLESS and NAC methods. Stars indicate *p*-values less than 0.05.

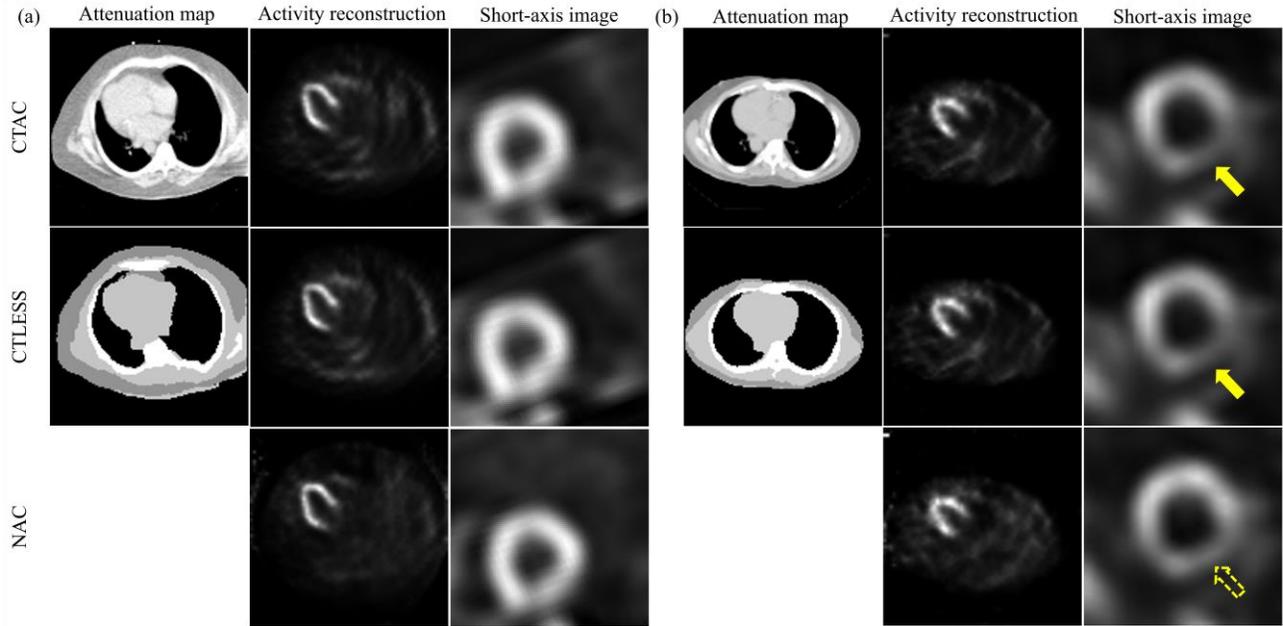

Fig. 12. Examples of SPECT images and attenuation maps obtained by CTLESS, CTAC, and NAC methods: (a) a defect-absent example, (b) a defect-present example, where yellow arrows indicate the defect. Red arrows indicate false defects introduced when the NAC method was used.

### C. Evaluation with different sizes of training dataset

Fig. 13a shows the AUC values with 95% CIs obtained using the CTLESS method and the EST-AC method. We observe that, even as the size of the training dataset reduced, the performance of CTLESS remained stable. Further, the AUC values yielded by CTLESS were higher than those yielded by the EST-AC method in the mean sense. Fig. 13b and Fig. 13c show RMSE and SSIM between the CTAC- derived activity maps and those obtained by the CTLESS method and the EST-AC method. For all sizes of the training dataset, we observed that the CTLESS method statistically outperformed the EST-AC method in terms of RMSE and SSIM ($p$-values < 0.05).

## IV. DISCUSSION

In this study, we developed and objectively evaluated a method for AC in SPECT without requiring a transmission scan, namely, the CTLESS method. Evaluation results using retrospective studies with anonymized clinical data and a model observer showed that CTLESS was statistically non-inferior to the CTAC method within a margin of 5% AUC obtained by the CTAC method and significantly outperformed the NAC method on the cardiac perfusion defect-detection task. Qualitatively identical results are found when the performance data are analyzed as precision-recall curves (not shown). In addition, the CTLESS method yielded visually similar attenuation maps as derived from the CT scans and similar reconstructed activity maps to those obtained by CTAC in both defect-present samples and defect-absent samples.

A key goal of our evaluation study was to assess the performance of the CTLESS method across a range of clinically relevant patient populations. In this context, coronary artery disease occurs in both male and female subjects who have very different body habitus [49], [50]. We observed that the CTLESS method yielded similar performance to the CTAC method and significantly outperformed the NAC method for both sexes.

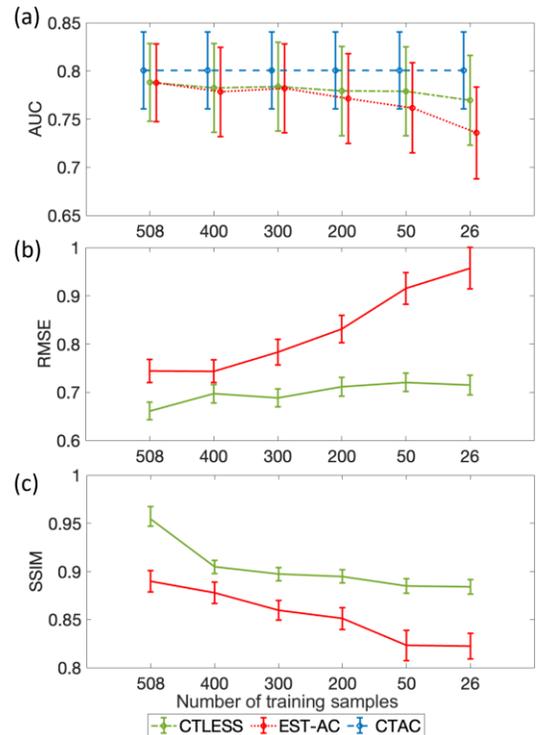

Fig. 13. AUC, RMSE, and SSIM obtained using the CTLESS method and the EST-AC method with different sizes of the training dataset.

Similarly, myocardial perfusion defects vary in severity, extent, and location in patients. These characteristics of defects are associated with coronary artery disease severity [51]. In the evaluation study, we considered 27 types of defects similar to previous studies. Our results showed that the CTLESS method yielded similar performance to the CTAC method for all defect severity and extents, indicating the robustness of the method across different defect types.

Clinical adoption of DL-based medical-imaging methods requires that they generalize across different clinical scanners [29]. This is very important, because otherwise the method will have to be trained using data acquired from each scanner, making the method impractical. In our evaluation study, we observed that the CTLESS method generalized across two different scanners, and more specifically, yielded equivalent performance as the CTAC method when trained on one scanner and tested on another. Notably, while from the same vendor, these scanners had different detectors, one a more traditional NaI detector and another a more advanced CZT detector. This finding is encouraging and motivates further research to assess whether the CTLESS method could be trained on data from an existing clinical scanner but then be applied to data from newer scanners, including those from different vendors. Another notable feature of our study is that we evaluated the generalizability of the method on the clinical task of interest, making these findings encouraging for clinical relevance and impact.

Another important consideration for the clinical adoption of AI-based methods for medical imaging is the requirement for training data. In this context, we observe that for multiple different sizes of the training dataset, the CTLESS method was not impacted as the training dataset size was reduced. Further, CTLESS outperformed another approach to determine the attenuation map that estimates the attenuation coefficient of each voxel (Fig. 13). To explain this finding, we recognize that since the relation between SPECT emission data and attenuation coefficients for each voxel can be different and complex, a DL-based approach can require large amounts of training data to estimate attenuation coefficients for all the voxels directly. The proposed CTLESS method alleviates this issue by recognizing that the attenuation coefficient is almost constant within the same organ in the torso region. Further, the attenuation coefficients of the different organs are approximately known. The CTLESS method uses this prior information to posit the determination of the attenuation map as a segmentation problem instead of estimating the attenuation coefficient for each voxel. Incorporating this prior information reduces the space of potential solutions, helping reduce the requirement for large training data. A tradeoff with the CTLESS method though is that assuming known attenuation coefficients can be inaccurate in organs such as lungs where the density varies depending on several factors, including disease state [52], [53]. However, in our data acquisition protocol, the SPECT images were acquired over 180° ranging from 45° right anterior oblique to 45° left posterior oblique, so the lung attenuation was less relevant. Nevertheless, it shows that evaluating the CTLESS method with clinical data is crucial, forming one of the reasons for our retrospective evaluation study.

The proposed CTLESS method could offer a direct advantage to MPI studies conducted on SPECT-only scanners, as it enables AC without CT images. As noted earlier, a substantial fraction of SPECT systems lack CT components. The ability to perform CT-less AC in those systems can make AC MPI studies available to larger numbers of patients. In addition, the CTLESS method inherently avoids misalignment between SPECT and CT scans, thus avoiding the potential for inaccurate diagnosis in MPI that may occur due to this misalignment. A recent study observed that up to 42% of MPI studies had moderate to severe misalignment [54]. Such misalignment can lead to artifactual perfusion defects [55]. Even a misalignment of one pixel can cause artifacts in the anterior, apical, and septal segments. A misalignment of 2-3 cm can cause a 20%- 35% change in apparent myocardial activity [56]. CTLESS is not sensitive to this misalignment since any patient motion that occurs during the scan should equally affect the photopeak and the scatter window data. To further assess this effect, we considered cases in our dataset where the CT and SPECT scans were misaligned and compared reconstructions obtained by CTLESS and CTAC methods. Our analysis indicated that while there were cases where false defects appeared when the SPECT and CT were misaligned, the output obtained by the CTLESS method had fewer such occurrences. A representative example is shown in Fig. 14, where we observed that there was a false-positive defect in the anterior LV wall when SPECT and CT were misaligned. However, in this example, we did not observe this false-positive defect in the images obtained using CTLESS as well as in the images when SPECT and CT were well registered. Similar findings were observed in another study where misalignments between SPECT and CT scans were deliberately introduced to study this effect [57]. Indeed, CTLESS also had imperfect estimates of attenuation maps. However, our initial observations indicated that, on the defect-detection task, the effect of this error was relatively smaller compared to the SPECT/CT misalignment due to patient motion. Other advantages of CTLESS include the lower radiation dose and acquisition costs. Abdollahi et al. observed that acquiring low-dose CT images in MPI study for AC purpose increased the effective dose by around 5% [58]. The as low as reasonably achievable (ALARA) principle promotes making every reasonable effort to maintain exposures to ionizing radiation as low as practical, considering multiple considerations, including the state of technology [59]. Thus, unless CT acquisition is clinically necessary, such as when intended for diagnostic purposes, the ALARA principle would favor the CTLESS method, a non-inferior imaging method that eliminates the exposure caused by CT scans.

In this study, the input to CTLESS were images reconstructed in scatter-window and photopeak-window sinogram data in binned format. Studies have reported improved performance of methods on clinical tasks using list-mode data, compared with data in binned format in SPECT [60]–[64]. Of most relevance, Rahman et al. have quantitatively shown that list-mode data contains more information than binned data for estimating attenuation coefficients [22]. Therefore, advancing the CTLESS method for list-mode data may yield even better performance. Moreover, in the CTLESS method, the initial estimate of the attenuation map was reconstructed using an OSEM-based approach that did not model the entire physics of the scatter-window data. However, recently, methods have been developed towards performing such inversion to estimate the attenuation map [15], [65]. While these methods have yielded limited performance so far, as they continue to advance, they could be used to compute the initial estimate of the attenuation map. Another area of research is directly using the scatter-window data as input for a DL method for AC [66] and comparing the performance to the proposed approach.

In addition to the above future directions, this study has some other limitations. First, in our evaluation studies, we used images with synthetic defects. Ideally, CTLESS should be evaluated using data with real perfusion defects. While the clinical records could provide knowledge of the presence of the defect, the accuracy of diagnosis and location of the defect could not be determined reliably due to inter- and intra-reader variability. One way to overcome this limitation is the use of cardiac catheterization to obtain a surrogate ground truth for the defect. While such a study would be time-consuming and expensive, our results motivate such an evaluation. Second, we used a model observer study to evaluate the defect-detection performance of the CTLESS method. While this observer has been shown to mimic human observer performance in MPI SPECT studies, ideally, the study should be conducted using experienced human observers. The results from the model observer study motivate the evaluation of the method using experienced human observers. A third limitation is that the performance of CTLESS method relies on the quality of its training data, particularly the ground-truth segmentation of CT images. In our approach, we employed a Markov random field-based method for CT image segmentation. Alternative techniques that are specifically designed for low-dose CT segmentation may provide better ground-truth segmentation masks. In our study, we have a total of N = 508 MPI studies in the training dataset and N = 266 MPI studies in the test dataset. To increase the robustness and generalization of CTLESS method, possible data augmentation strategies could be considered [67]–[69]. In addition, previous studies have shown reduced diagnostic accuracy due to attenuation artifacts in overweight and obese patients in MPI-SPECT studies [70]–[72]. Thus, a stratified analysis of CTLESS for overweight and obese patients is another important research direction. Moreover, the study was conducted with data acquired at a single center and with SPECT scanners from a single vendor. However, the results of this study motivate evaluation studies with data from other scanners and centers.

## V. Conclusion

The proposed scatter-window projection and deep learning-based transmission-less attenuation compensation (AC) method (CTLESS) for myocardial perfusion SPECT yielded statistically non-inferior performance compared to a standard CT-based AC method (CTAC) on the clinically relevant task of detecting myocardial perfusion defects, as evaluated in a retrospective clinical study in patients who underwent myocardial perfusion imaging SPECT studies. The method yielded a similar performance as the CTAC method for different defect extents and severities as well as for different patient sexes. In addition, the method was observed to generalize across two SPECT scanners. Further, the method outperformed a method where no AC was performed on both the clinical task of detecting perfusion defects and using metrics that quantify visual fidelity. As per the RELAINCE guidelines, we derive the following claim for the proposed CTLESS method:

*A deep learning-based transmission-less AC method for myocardial perfusion SPECT yielded statistically non-inferior performance to a standard CT-based AC method on the task of detecting myocardial perfusion defects within a margin of 5% of AUC with a significance level of 0.05 as evaluated in a retrospective study with single-center multi-scanner data and with an anthropomorphic model observer.*

Software to conduct this study and supplementary materials are available at https://drive.google.com/drive/folders/1HYzDyWTXes5oCN-ONxEeCnuxgkZ9y2eD?usp=drive_link .